\documentstyle[12pt]{article}
\begin{document}

\pagestyle{empty}
\vspace*{3cm}
\begin{center}
{\Large\bf MULTIPLICITY MOMENTS IN QCD AND EXPERIMENT}

\bigskip

\bigskip

{\large I.M. Dremin}

{\it P.N. Lebedev Physical Institute, Moscow, Russia}
\end{center}

\vspace{5cm}

\begin{abstract}
QCD predictions for moments of {\it parton} multiplicity distributions
are discussed. The next-to-leading terms and conservation law give rise
to the peculiar oscillating shape of some ratio of the moments. The
similar shape has been found by moment analysis of {\it hadron}
multiplicities. Experimental data, theoretical Monte Carlo models and
phenomenological fits have been used for ee, hh, hA, AA reactions at
high energies.
\end{abstract}

\newpage

\baselineskip 21.75pt 

In multiparticle production the primary observable is the multiplicity
distribution $P_n$ which contains in the integrated form all the
correlations of the system. First attempts to derive {\it parton}
multiplicities in the double logarithmic approximation (DLA) of QCD
provided extremely wide distributions compared to those found
experimentally for {\it hadrons}. Later, the solution of the QCD
equation for the generating function which takes into account higher
order terms of the perturbative expansion and conservation law was
found \cite{1} and more efficient way of studying the distribution shape
based on moment analysis was proposed (see review \cite{2}). When
applied to {\it hadron} multiplicities, it shows peculiar oscillating
curves first derived for {\it partons} in QCD \cite{3}. Their stability
for reactions initiated by different projectiles and targets is rather
astonishing \cite{4}.

Let us describe briefly the theoretical findings. The generating
function is defined by
\begin{equation}
G(z)=\sum _{n=0}^{\infty }z^{n}P_n  \label{1}
\end{equation}
so that probabilities $P_n$, factorial ($F_q$) and cumulant ($K_q$)
moments are
\begin{equation}
P_n = \left. \frac {1}{n!}\frac {d^n G(z)}{dz^n}\right| _{z=0} , \,\,
F_q = \left. \frac {1}{\langle n\rangle ^q}\frac {d^q G(z)}{dz^q}\right|
_{z=1} ,\,\, K_q = \left. \frac {1}{\langle n\rangle ^q}\frac {d^q \ln
G(z)}{dz^q}\right| _{z=1} \label{2}
\end{equation}
with the recurrence relations
\begin{equation}
F_q = \sum _{m=0}^{q-1}C_{q-1}^{m}K_{q-m}F_m , \label{3}
\end{equation}
where $C_{n}^{m}=n!/m!(n-m)!$ are the binomial coefficients.

The equation for the generating function in gluodynamics \cite {5} looks
 like an ordinary birth-death equation but with a typical singular QCD
kernel
\begin{equation}
\frac {dG}{dy}=\int _{0}^{1}dxK(x)\gamma _{0}^{2}[G(y+\ln (1-x))G(y+\ln
x) - G(y)],  \label{4}
\end{equation}
where $\gamma _{0}^{2}=2N_{c}\alpha _{s}/\pi $, $\alpha _{s}$ is the
coupling constant, $N_c =3$ is the number of colours, $K(x)=\frac
{1}{x}-(1-x)[2-x(1-x)], \, y=\ln Q^2 /Q_{0}^{2}, \, Q^2$ is the jet
virtuality, $Q_{0}^{2}$=const. The $y$-dependence enters $G(z,y)$
through the energy dependence of $P_n$. The variable $z$ is omitted in
eq.(\ref{4}).

For fixed coupling constant, this equation can be reduced to the
algebraic one and, therefore, has an exact solution
\cite{6, 7}. For running coupling, one can use Taylor series expansion
of the terms in the non-linear part of the integral.
It is instructive to demonstrate it here because lowest order terms
give rise to DLA, and higher orders provide modified leading logarithm
and next-to-leading logarithm approximations (see \cite{1,2}).
After the expansion of $G$ at large $y$ in non-linear terms is done and
both sides of eq.(\ref{4}) are divided by $G(y)$, one differentiates
both sides of (\ref{4}) and the non-linear integro-differential equation
(\ref{4}) is reduced to the differential equation
\begin{equation}
(\ln G)'' = \gamma _{0}[G-1-2h_{1}G'+h_{2}G''], \label{5}
\end{equation}
where $h_1 =11/24, h_2 \approx 0.216$ are given by integrals of the
kernel $K(x)$ with corresponding weights. The various approximations are
obtained by considering in the righthand side of (\ref{5}) only two
(DLA), three (MLLA) or four (NLLA) terms. Higher order derivatives are
omitted here.

Using the relation (\ref{2}) and equating the terms of the same power of
$z$ on both sides of the equation, one gets from (\ref{5})
for the ratio of cumulant to factorial moments
\begin{equation}
H_q \equiv \frac {K_q}{F_q}=\frac {\gamma _{0}^{2}[1-2h_{1}q\gamma +
h_{2}(q^{2}\gamma ^{2}+q\gamma ')]}{q^{2}\gamma ^{2}+q\gamma '}
\label{6}
\end{equation}
with
\begin{equation}
\gamma \approx \gamma _{0}-\frac {1}{2}h_{1}\gamma _{0}^{2}+\frac
{1}{8}(4h_{2}-h_{1}^{2})\gamma _{0}^{3}. \label{7}
\end{equation}
In DLA, one neglects all the terms of higher order in $\gamma _0$
 and gets from (\ref{6})
\begin{equation}
H_{q}^{(DLA)}=\frac {1}{q^2}. \label{8}
\end{equation}
Let us note that such a monotonous behaviour of the ever positive $H_q$
corresponds to the negative
binomial distribution with its parameter $k=2$. The experimental fits
show much larger values of $k$ and, consequently, more narrow
distributions.

For the realistic values of $\gamma _{0}$=0.48, the ratio $H_q$ as given
by eqs. (\ref{5}), (\ref{6})
exhibits the minimum at
\begin{equation}
q_{min} \approx \frac {1}{h_{1}\gamma _{0}}\approx 5 \label{9}
\end{equation}
and then increases. With higher order terms of Taylor series expansion
taken into account the ratio $H_q$ oscillates \cite{3}.

These oscillations are due to oscillatory behaviour of cumulants which
have the physical meaning of genuine correlations in the $q$-parton
system. The location of the first minimum is determined by the condition
$q\gamma \approx 1$ which is related to the well-known quantum-field
theory problem of the break-up of the perturbative expansion due to the
factorial increase of the number of graphs for $n$-particle process.
In traditional language, it corresponds to
the expansion parameter $\alpha _{s} n$ appearing instead of $\alpha _s$
in that case. At the
same time, this location is determined by the singular nature of the
kernel at $x\rightarrow 0 $ (and $x\rightarrow 1$ in the symmetrized
form). For theories with regular kernels like $\lambda \phi _{6}^{3}$
\cite{8} the first minimum is located at larger values of $q$.

The KNO-scaling is valid in such an approach, in practice. The energy
dependence of $H_q$ should be extremely weak. It is only due to running
property of $\gamma _0$ and almost cancels in the ratio $H_q$ as seen
from (\ref{6}).

Surely, there is no experimental information on parton distributions.
The local parton-hadron duality is often used to relate them to hadron
distributions by imposing proportionality assumption. Otherwise, the
hadronization scheme should be developed. Anyway, one may not confront
directly theoretical predictions about partons to experimental data.

Nevertheless, QCD shows a way to the new method of analysis of experimental
data. Cumulants are very sensitive to slight variations of the
distributions because of subtractions implied in eq.(\ref{3}).
Therefore,
 it is worthwhile to calculate factorial moments of experimental
multiplicity distributions and then
cumulants, using (\ref{3}), to find out the behaviour of the ratio $H_q$
for hadrons. Its values in $e^{+}e^{-}$ collisions at $Z^0$ peak
\cite{9,10} and in hadronic reactions \cite{4, 9} are shown in Figs.1 and
2, correspondingly.

\begin{figure}
\setlength{\unitlength}{0.240900pt}
\ifx\plotpoint\undefined\newsavebox{\plotpoint}\fi
\sbox{\plotpoint}{\rule[-0.200pt]{0.400pt}{0.400pt}}%
\begin{picture}(1500,900)(0,0)
\font\gnuplot=cmr10 at 10pt
\gnuplot
\sbox{\plotpoint}{\rule[-0.200pt]{0.400pt}{0.400pt}}%
\put(220.0,266.0){\rule[-0.200pt]{292.934pt}{0.400pt}}
\put(220.0,113.0){\rule[-0.200pt]{4.818pt}{0.400pt}}
\put(198,113){\makebox(0,0)[r]{-0.002}}
\put(1416.0,113.0){\rule[-0.200pt]{4.818pt}{0.400pt}}
\put(220.0,189.0){\rule[-0.200pt]{4.818pt}{0.400pt}}
\put(198,189){\makebox(0,0)[r]{-0.001}}
\put(1416.0,189.0){\rule[-0.200pt]{4.818pt}{0.400pt}}
\put(220.0,266.0){\rule[-0.200pt]{4.818pt}{0.400pt}}
\put(198,266){\makebox(0,0)[r]{0}}
\put(1416.0,266.0){\rule[-0.200pt]{4.818pt}{0.400pt}}
\put(220.0,342.0){\rule[-0.200pt]{4.818pt}{0.400pt}}
\put(198,342){\makebox(0,0)[r]{0.001}}
\put(1416.0,342.0){\rule[-0.200pt]{4.818pt}{0.400pt}}
\put(220.0,419.0){\rule[-0.200pt]{4.818pt}{0.400pt}}
\put(198,419){\makebox(0,0)[r]{0.002}}
\put(1416.0,419.0){\rule[-0.200pt]{4.818pt}{0.400pt}}
\put(220.0,495.0){\rule[-0.200pt]{4.818pt}{0.400pt}}
\put(198,495){\makebox(0,0)[r]{0.003}}
\put(1416.0,495.0){\rule[-0.200pt]{4.818pt}{0.400pt}}
\put(220.0,571.0){\rule[-0.200pt]{4.818pt}{0.400pt}}
\put(198,571){\makebox(0,0)[r]{0.004}}
\put(1416.0,571.0){\rule[-0.200pt]{4.818pt}{0.400pt}}
\put(220.0,648.0){\rule[-0.200pt]{4.818pt}{0.400pt}}
\put(198,648){\makebox(0,0)[r]{0.005}}
\put(1416.0,648.0){\rule[-0.200pt]{4.818pt}{0.400pt}}
\put(220.0,724.0){\rule[-0.200pt]{4.818pt}{0.400pt}}
\put(198,724){\makebox(0,0)[r]{0.006}}
\put(1416.0,724.0){\rule[-0.200pt]{4.818pt}{0.400pt}}
\put(220.0,801.0){\rule[-0.200pt]{4.818pt}{0.400pt}}
\put(198,801){\makebox(0,0)[r]{0.007}}
\put(1416.0,801.0){\rule[-0.200pt]{4.818pt}{0.400pt}}
\put(220.0,877.0){\rule[-0.200pt]{4.818pt}{0.400pt}}
\put(198,877){\makebox(0,0)[r]{0.008}}
\put(1416.0,877.0){\rule[-0.200pt]{4.818pt}{0.400pt}}
\put(220.0,113.0){\rule[-0.200pt]{0.400pt}{4.818pt}}
\put(220,68){\makebox(0,0){2}}
\put(220.0,857.0){\rule[-0.200pt]{0.400pt}{4.818pt}}
\put(382.0,113.0){\rule[-0.200pt]{0.400pt}{4.818pt}}
\put(382,68){\makebox(0,0){4}}
\put(382.0,857.0){\rule[-0.200pt]{0.400pt}{4.818pt}}
\put(544.0,113.0){\rule[-0.200pt]{0.400pt}{4.818pt}}
\put(544,68){\makebox(0,0){6}}
\put(544.0,857.0){\rule[-0.200pt]{0.400pt}{4.818pt}}
\put(706.0,113.0){\rule[-0.200pt]{0.400pt}{4.818pt}}
\put(706,68){\makebox(0,0){8}}
\put(706.0,857.0){\rule[-0.200pt]{0.400pt}{4.818pt}}
\put(869.0,113.0){\rule[-0.200pt]{0.400pt}{4.818pt}}
\put(869,68){\makebox(0,0){10}}
\put(869.0,857.0){\rule[-0.200pt]{0.400pt}{4.818pt}}
\put(1031.0,113.0){\rule[-0.200pt]{0.400pt}{4.818pt}}
\put(1031,68){\makebox(0,0){12}}
\put(1031.0,857.0){\rule[-0.200pt]{0.400pt}{4.818pt}}
\put(1193.0,113.0){\rule[-0.200pt]{0.400pt}{4.818pt}}
\put(1193,68){\makebox(0,0){14}}
\put(1193.0,857.0){\rule[-0.200pt]{0.400pt}{4.818pt}}
\put(1355.0,113.0){\rule[-0.200pt]{0.400pt}{4.818pt}}
\put(1355,68){\makebox(0,0){16}}
\put(1355.0,857.0){\rule[-0.200pt]{0.400pt}{4.818pt}}
\put(220.0,113.0){\rule[-0.200pt]{292.934pt}{0.400pt}}
\put(1436.0,113.0){\rule[-0.200pt]{0.400pt}{184.048pt}}
\put(220.0,877.0){\rule[-0.200pt]{292.934pt}{0.400pt}}
\put(59,405){\makebox(0,0){$H_q$}}
\put(828,23){\makebox(0,0){$q$}}
\put(220.0,113.0){\rule[-0.200pt]{0.400pt}{184.048pt}}
\put(1306,812){\makebox(0,0)[r]{$e^+e^-$}}
\put(1328.0,812.0){\rule[-0.200pt]{15.899pt}{0.400pt}}
\put(301,801){\usebox{\plotpoint}}
\multiput(301.58,789.93)(0.499,-3.221){159}{\rule{0.120pt}{2.668pt}}
\multiput(300.17,795.46)(81.000,-514.463){2}{\rule{0.400pt}{1.334pt}}
\multiput(382.00,279.92)(0.533,-0.499){149}{\rule{0.526pt}{0.120pt}}
\multiput(382.00,280.17)(79.908,-76.000){2}{\rule{0.263pt}{0.400pt}}
\multiput(463.00,205.58)(1.779,0.496){43}{\rule{1.509pt}{0.120pt}}
\multiput(463.00,204.17)(77.869,23.000){2}{\rule{0.754pt}{0.400pt}}
\multiput(544.00,228.58)(1.359,0.497){57}{\rule{1.180pt}{0.120pt}}
\multiput(544.00,227.17)(78.551,30.000){2}{\rule{0.590pt}{0.400pt}}
\multiput(625.00,258.58)(1.779,0.496){43}{\rule{1.509pt}{0.120pt}}
\multiput(625.00,257.17)(77.869,23.000){2}{\rule{0.754pt}{0.400pt}}
\put(706,279.67){\rule{19.513pt}{0.400pt}}
\multiput(706.00,280.17)(40.500,-1.000){2}{\rule{9.756pt}{0.400pt}}
\multiput(787.00,278.92)(4.248,-0.491){17}{\rule{3.380pt}{0.118pt}}
\multiput(787.00,279.17)(74.985,-10.000){2}{\rule{1.690pt}{0.400pt}}
\multiput(869.00,268.92)(3.469,-0.492){21}{\rule{2.800pt}{0.119pt}}
\multiput(869.00,269.17)(75.188,-12.000){2}{\rule{1.400pt}{0.400pt}}
\multiput(950.00,256.93)(8.948,-0.477){7}{\rule{6.580pt}{0.115pt}}
\multiput(950.00,257.17)(67.343,-5.000){2}{\rule{3.290pt}{0.400pt}}
\multiput(1031.00,253.59)(4.689,0.489){15}{\rule{3.700pt}{0.118pt}}
\multiput(1031.00,252.17)(73.320,9.000){2}{\rule{1.850pt}{0.400pt}}
\multiput(1112.00,262.58)(2.285,0.495){33}{\rule{1.900pt}{0.119pt}}
\multiput(1112.00,261.17)(77.056,18.000){2}{\rule{0.950pt}{0.400pt}}
\multiput(1193.00,280.61)(17.877,0.447){3}{\rule{10.900pt}{0.108pt}}
\multiput(1193.00,279.17)(58.377,3.000){2}{\rule{5.450pt}{0.400pt}}
\multiput(1274.00,281.92)(2.755,-0.494){27}{\rule{2.260pt}{0.119pt}}
\multiput(1274.00,282.17)(76.309,-15.000){2}{\rule{1.130pt}{0.400pt}}
\put(1350,812){\raisebox{-.8pt}{\makebox(0,0){$\Box$}}}
\put(301,801){\raisebox{-.8pt}{\makebox(0,0){$\Box$}}}
\put(382,281){\raisebox{-.8pt}{\makebox(0,0){$\Box$}}}
\put(463,205){\raisebox{-.8pt}{\makebox(0,0){$\Box$}}}
\put(544,228){\raisebox{-.8pt}{\makebox(0,0){$\Box$}}}
\put(625,258){\raisebox{-.8pt}{\makebox(0,0){$\Box$}}}
\put(706,281){\raisebox{-.8pt}{\makebox(0,0){$\Box$}}}
\put(787,280){\raisebox{-.8pt}{\makebox(0,0){$\Box$}}}
\put(869,270){\raisebox{-.8pt}{\makebox(0,0){$\Box$}}}
\put(950,258){\raisebox{-.8pt}{\makebox(0,0){$\Box$}}}
\put(1031,253){\raisebox{-.8pt}{\makebox(0,0){$\Box$}}}
\put(1112,262){\raisebox{-.8pt}{\makebox(0,0){$\Box$}}}
\put(1193,280){\raisebox{-.8pt}{\makebox(0,0){$\Box$}}}
\put(1274,283){\raisebox{-.8pt}{\makebox(0,0){$\Box$}}}
\put(1355,268){\raisebox{-.8pt}{\makebox(0,0){$\Box$}}}
\end{picture}

\caption{}
\end{figure}

\begin{figure}
\setlength{\unitlength}{0.240900pt}
\ifx\plotpoint\undefined\newsavebox{\plotpoint}\fi
\sbox{\plotpoint}{\rule[-0.200pt]{0.400pt}{0.400pt}}%
\begin{picture}(1500,900)(0,0)
\font\gnuplot=cmr10 at 10pt
\gnuplot
\sbox{\plotpoint}{\rule[-0.200pt]{0.400pt}{0.400pt}}%
\put(220.0,419.0){\rule[-0.200pt]{292.934pt}{0.400pt}}
\put(220.0,113.0){\rule[-0.200pt]{4.818pt}{0.400pt}}
\put(198,113){\makebox(0,0)[r]{-0.1}}
\put(1416.0,113.0){\rule[-0.200pt]{4.818pt}{0.400pt}}
\put(220.0,266.0){\rule[-0.200pt]{4.818pt}{0.400pt}}
\put(198,266){\makebox(0,0)[r]{-0.05}}
\put(1416.0,266.0){\rule[-0.200pt]{4.818pt}{0.400pt}}
\put(220.0,419.0){\rule[-0.200pt]{4.818pt}{0.400pt}}
\put(198,419){\makebox(0,0)[r]{0}}
\put(1416.0,419.0){\rule[-0.200pt]{4.818pt}{0.400pt}}
\put(220.0,571.0){\rule[-0.200pt]{4.818pt}{0.400pt}}
\put(198,571){\makebox(0,0)[r]{0.05}}
\put(1416.0,571.0){\rule[-0.200pt]{4.818pt}{0.400pt}}
\put(220.0,724.0){\rule[-0.200pt]{4.818pt}{0.400pt}}
\put(198,724){\makebox(0,0)[r]{0.1}}
\put(1416.0,724.0){\rule[-0.200pt]{4.818pt}{0.400pt}}
\put(220.0,877.0){\rule[-0.200pt]{4.818pt}{0.400pt}}
\put(198,877){\makebox(0,0)[r]{0.15}}
\put(1416.0,877.0){\rule[-0.200pt]{4.818pt}{0.400pt}}
\put(220.0,113.0){\rule[-0.200pt]{0.400pt}{4.818pt}}
\put(220,68){\makebox(0,0){3}}
\put(220.0,857.0){\rule[-0.200pt]{0.400pt}{4.818pt}}
\put(372.0,113.0){\rule[-0.200pt]{0.400pt}{4.818pt}}
\put(372,68){\makebox(0,0){4}}
\put(372.0,857.0){\rule[-0.200pt]{0.400pt}{4.818pt}}
\put(524.0,113.0){\rule[-0.200pt]{0.400pt}{4.818pt}}
\put(524,68){\makebox(0,0){5}}
\put(524.0,857.0){\rule[-0.200pt]{0.400pt}{4.818pt}}
\put(676.0,113.0){\rule[-0.200pt]{0.400pt}{4.818pt}}
\put(676,68){\makebox(0,0){6}}
\put(676.0,857.0){\rule[-0.200pt]{0.400pt}{4.818pt}}
\put(828.0,113.0){\rule[-0.200pt]{0.400pt}{4.818pt}}
\put(828,68){\makebox(0,0){7}}
\put(828.0,857.0){\rule[-0.200pt]{0.400pt}{4.818pt}}
\put(980.0,113.0){\rule[-0.200pt]{0.400pt}{4.818pt}}
\put(980,68){\makebox(0,0){8}}
\put(980.0,857.0){\rule[-0.200pt]{0.400pt}{4.818pt}}
\put(1132.0,113.0){\rule[-0.200pt]{0.400pt}{4.818pt}}
\put(1132,68){\makebox(0,0){9}}
\put(1132.0,857.0){\rule[-0.200pt]{0.400pt}{4.818pt}}
\put(1284.0,113.0){\rule[-0.200pt]{0.400pt}{4.818pt}}
\put(1284,68){\makebox(0,0){10}}
\put(1284.0,857.0){\rule[-0.200pt]{0.400pt}{4.818pt}}
\put(1436.0,113.0){\rule[-0.200pt]{0.400pt}{4.818pt}}
\put(1436,68){\makebox(0,0){11}}
\put(1436.0,857.0){\rule[-0.200pt]{0.400pt}{4.818pt}}
\put(220.0,113.0){\rule[-0.200pt]{292.934pt}{0.400pt}}
\put(1436.0,113.0){\rule[-0.200pt]{0.400pt}{184.048pt}}
\put(220.0,877.0){\rule[-0.200pt]{292.934pt}{0.400pt}}
\put(89,405){\makebox(0,0){$H_q$}}
\put(828,23){\makebox(0,0){q}}
\put(220.0,113.0){\rule[-0.200pt]{0.400pt}{184.048pt}}
\put(1306,812){\makebox(0,0)[r]{$^{32}$S $^{238}$ U, DPM}}
\put(1328.0,812.0){\rule[-0.200pt]{15.899pt}{0.400pt}}
\put(220,798){\usebox{\plotpoint}}
\multiput(220.58,791.59)(0.499,-1.809){301}{\rule{0.120pt}{1.545pt}}
\multiput(219.17,794.79)(152.000,-545.794){2}{\rule{0.400pt}{0.772pt}}
\multiput(372.00,247.92)(0.571,-0.499){263}{\rule{0.557pt}{0.120pt}}
\multiput(372.00,248.17)(150.844,-133.000){2}{\rule{0.279pt}{0.400pt}}
\multiput(524.58,116.00)(0.499,0.892){301}{\rule{0.120pt}{0.813pt}}
\multiput(523.17,116.00)(152.000,269.312){2}{\rule{0.400pt}{0.407pt}}
\multiput(676.58,387.00)(0.499,1.387){301}{\rule{0.120pt}{1.208pt}}
\multiput(675.17,387.00)(152.000,418.493){2}{\rule{0.400pt}{0.604pt}}
\multiput(828.00,806.93)(16.851,-0.477){7}{\rule{12.260pt}{0.115pt}}
\multiput(828.00,807.17)(126.554,-5.000){2}{\rule{6.130pt}{0.400pt}}
\multiput(980.58,791.67)(0.499,-3.295){207}{\rule{0.120pt}{2.729pt}}
\multiput(979.17,797.34)(105.000,-684.337){2}{\rule{0.400pt}{1.364pt}}
\multiput(1395.58,113.00)(0.498,7.142){79}{\rule{0.120pt}{5.768pt}}
\multiput(1394.17,113.00)(41.000,569.028){2}{\rule{0.400pt}{2.884pt}}
\put(1436,694){\usebox{\plotpoint}}
\put(1350,812){\makebox(0,0){$+$}}
\put(220,798){\makebox(0,0){$+$}}
\put(372,249){\makebox(0,0){$+$}}
\put(524,116){\makebox(0,0){$+$}}
\put(676,387){\makebox(0,0){$+$}}
\put(828,808){\makebox(0,0){$+$}}
\put(980,803){\makebox(0,0){$+$}}
\put(1436,694){\makebox(0,0){$+$}}
\sbox{\plotpoint}{\rule[-0.500pt]{1.000pt}{1.000pt}}%
\put(1306,767){\makebox(0,0)[r]{p Al, DPM}}
\multiput(1328,767)(41.511,0.000){2}{\usebox{\plotpoint}}
\put(1394,767){\usebox{\plotpoint}}
\put(220,834){\usebox{\plotpoint}}
\multiput(220,834)(19.064,-36.874){8}{\usebox{\plotpoint}}
\multiput(372,540)(30.135,-28.549){6}{\usebox{\plotpoint}}
\multiput(524,396)(40.333,-9.818){3}{\usebox{\plotpoint}}
\multiput(676,359)(40.871,7.260){4}{\usebox{\plotpoint}}
\multiput(828,386)(39.874,11.542){4}{\usebox{\plotpoint}}
\multiput(980,430)(40.917,6.999){4}{\usebox{\plotpoint}}
\multiput(1132,456)(41.421,-2.725){3}{\usebox{\plotpoint}}
\multiput(1284,446)(40.333,-9.818){4}{\usebox{\plotpoint}}
\put(1436,409){\usebox{\plotpoint}}
\sbox{\plotpoint}{\rule[-0.400pt]{0.800pt}{0.800pt}}%
\put(1306,722){\makebox(0,0)[r]{p Al, QGSM}}
\put(1328.0,722.0){\rule[-0.400pt]{15.899pt}{0.800pt}}
\put(220,647){\usebox{\plotpoint}}
\multiput(221.41,640.82)(0.501,-0.807){297}{\rule{0.121pt}{1.489pt}}
\multiput(218.34,643.91)(152.000,-241.909){2}{\rule{0.800pt}{0.745pt}}
\multiput(372.00,400.09)(0.965,-0.501){151}{\rule{1.739pt}{0.121pt}}
\multiput(372.00,400.34)(148.390,-79.000){2}{\rule{0.870pt}{0.800pt}}
\multiput(524.00,324.41)(2.084,0.503){67}{\rule{3.486pt}{0.121pt}}
\multiput(524.00,321.34)(144.764,37.000){2}{\rule{1.743pt}{0.800pt}}
\multiput(676.00,361.41)(0.828,0.501){177}{\rule{1.522pt}{0.121pt}}
\multiput(676.00,358.34)(148.842,92.000){2}{\rule{0.761pt}{0.800pt}}
\multiput(828.00,453.41)(1.175,0.501){123}{\rule{2.071pt}{0.121pt}}
\multiput(828.00,450.34)(147.702,65.000){2}{\rule{1.035pt}{0.800pt}}
\multiput(980.00,515.09)(1.565,-0.502){91}{\rule{2.682pt}{0.121pt}}
\multiput(980.00,515.34)(146.434,-49.000){2}{\rule{1.341pt}{0.800pt}}
\multiput(1133.41,463.17)(0.501,-0.602){297}{\rule{0.121pt}{1.163pt}}
\multiput(1130.34,465.59)(152.000,-180.586){2}{\rule{0.800pt}{0.582pt}}
\multiput(1285.41,280.65)(0.501,-0.529){297}{\rule{0.121pt}{1.047pt}}
\multiput(1282.34,282.83)(152.000,-158.826){2}{\rule{0.800pt}{0.524pt}}
\put(1436,124){\usebox{\plotpoint}}
\sbox{\plotpoint}{\rule[-0.200pt]{0.400pt}{0.400pt}}%
\put(1306,677){\makebox(0,0)[r]{K Al, QGSM}}
\put(1328.0,677.0){\rule[-0.200pt]{15.899pt}{0.400pt}}
\put(220,492){\usebox{\plotpoint}}
\multiput(220.00,490.92)(0.655,-0.499){229}{\rule{0.624pt}{0.120pt}}
\multiput(220.00,491.17)(150.705,-116.000){2}{\rule{0.312pt}{0.400pt}}
\multiput(372.00,374.93)(11.553,-0.485){11}{\rule{8.786pt}{0.117pt}}
\multiput(372.00,375.17)(133.765,-7.000){2}{\rule{4.393pt}{0.400pt}}
\multiput(524.00,369.58)(1.736,0.498){85}{\rule{1.482pt}{0.120pt}}
\multiput(524.00,368.17)(148.924,44.000){2}{\rule{0.741pt}{0.400pt}}
\multiput(676.00,413.58)(1.736,0.498){85}{\rule{1.482pt}{0.120pt}}
\multiput(676.00,412.17)(148.924,44.000){2}{\rule{0.741pt}{0.400pt}}
\multiput(828.00,455.94)(22.122,-0.468){5}{\rule{15.300pt}{0.113pt}}
\multiput(828.00,456.17)(120.244,-4.000){2}{\rule{7.650pt}{0.400pt}}
\multiput(980.00,451.92)(1.137,-0.499){131}{\rule{1.007pt}{0.120pt}}
\multiput(980.00,452.17)(149.909,-67.000){2}{\rule{0.504pt}{0.400pt}}
\multiput(1132.00,384.92)(1.120,-0.499){133}{\rule{0.994pt}{0.120pt}}
\multiput(1132.00,385.17)(149.937,-68.000){2}{\rule{0.497pt}{0.400pt}}
\multiput(1284.00,318.58)(0.875,0.499){171}{\rule{0.799pt}{0.120pt}}
\multiput(1284.00,317.17)(150.342,87.000){2}{\rule{0.399pt}{0.400pt}}
\put(1436,405){\usebox{\plotpoint}}
\put(1306,632){\makebox(0,0)[r]{$\pi$ Al, QGSM}}
\multiput(1328,632)(20.756,0.000){4}{\usebox{\plotpoint}}
\put(1394,632){\usebox{\plotpoint}}
\put(220,604){\usebox{\plotpoint}}
\multiput(220,604)(14.202,-15.136){11}{\usebox{\plotpoint}}
\multiput(372,442)(19.129,-8.054){8}{\usebox{\plotpoint}}
\multiput(524,378)(20.734,-0.955){7}{\usebox{\plotpoint}}
\multiput(676,371)(20.337,4.148){8}{\usebox{\plotpoint}}
\multiput(828,402)(19.972,5.650){8}{\usebox{\plotpoint}}
\multiput(980,445)(20.480,3.368){7}{\usebox{\plotpoint}}
\multiput(1132,470)(20.458,-3.499){7}{\usebox{\plotpoint}}
\multiput(1284,444)(18.014,-10.310){9}{\usebox{\plotpoint}}
\put(1436,357){\usebox{\plotpoint}}
\put(1350,632){\makebox(0,0){$\times$}}
\put(220,604){\makebox(0,0){$\times$}}
\put(372,442){\makebox(0,0){$\times$}}
\put(524,378){\makebox(0,0){$\times$}}
\put(676,371){\makebox(0,0){$\times$}}
\put(828,402){\makebox(0,0){$\times$}}
\put(980,445){\makebox(0,0){$\times$}}
\put(1132,470){\makebox(0,0){$\times$}}
\put(1284,444){\makebox(0,0){$\times$}}
\put(1436,357){\makebox(0,0){$\times$}}
\put(1306,587){\makebox(0,0)[r]{p$\bar{\rm{p}}$, exp}}
\put(1328.0,587.0){\rule[-0.200pt]{15.899pt}{0.400pt}}
\put(220,610){\usebox{\plotpoint}}
\multiput(220.58,607.83)(0.499,-0.529){301}{\rule{0.120pt}{0.524pt}}
\multiput(219.17,608.91)(152.000,-159.913){2}{\rule{0.400pt}{0.262pt}}
\multiput(372.00,447.92)(1.387,-0.499){107}{\rule{1.205pt}{0.120pt}}
\multiput(372.00,448.17)(149.498,-55.000){2}{\rule{0.603pt}{0.400pt}}
\multiput(524.00,392.95)(33.728,-0.447){3}{\rule{20.367pt}{0.108pt}}
\multiput(524.00,393.17)(109.728,-3.000){2}{\rule{10.183pt}{0.400pt}}
\multiput(676.00,391.58)(3.673,0.496){39}{\rule{2.995pt}{0.119pt}}
\multiput(676.00,390.17)(145.783,21.000){2}{\rule{1.498pt}{0.400pt}}
\multiput(828.00,412.58)(4.299,0.495){33}{\rule{3.478pt}{0.119pt}}
\multiput(828.00,411.17)(144.782,18.000){2}{\rule{1.739pt}{0.400pt}}
\multiput(980.00,430.60)(22.122,0.468){5}{\rule{15.300pt}{0.113pt}}
\multiput(980.00,429.17)(120.244,4.000){2}{\rule{7.650pt}{0.400pt}}
\multiput(1132.00,432.92)(7.897,-0.491){17}{\rule{6.180pt}{0.118pt}}
\multiput(1132.00,433.17)(139.173,-10.000){2}{\rule{3.090pt}{0.400pt}}
\multiput(1284.00,422.92)(4.558,-0.495){31}{\rule{3.676pt}{0.119pt}}
\multiput(1284.00,423.17)(144.369,-17.000){2}{\rule{1.838pt}{0.400pt}}
\put(1436,407){\usebox{\plotpoint}}
\put(1350,587){\raisebox{-.8pt}{\makebox(0,0){$\Box$}}}
\put(220,610){\raisebox{-.8pt}{\makebox(0,0){$\Box$}}}
\put(372,449){\raisebox{-.8pt}{\makebox(0,0){$\Box$}}}
\put(524,394){\raisebox{-.8pt}{\makebox(0,0){$\Box$}}}
\put(676,391){\raisebox{-.8pt}{\makebox(0,0){$\Box$}}}
\put(828,412){\raisebox{-.8pt}{\makebox(0,0){$\Box$}}}
\put(980,430){\raisebox{-.8pt}{\makebox(0,0){$\Box$}}}
\put(1132,434){\raisebox{-.8pt}{\makebox(0,0){$\Box$}}}
\put(1284,424){\raisebox{-.8pt}{\makebox(0,0){$\Box$}}}
\put(1436,407){\raisebox{-.8pt}{\makebox(0,0){$\Box$}}}
\end{picture}
\caption{}
\end{figure}

One concludes that the oscillation pattern is observed in all the cases.
The position of the first minimum is in the range $q=4-6$. The value
of $H_q$ in it steadily increases (in modulus) from ee to pp, hA, AA.
The model calculations according to the dual parton model (DPM)
\cite{11} and to the quark-gluon string model (QGSM) \cite{12}
were used for hA and AA in Fig.2 but it is shown
in \cite{4} that they fit experimental data rather well. On the
contrary, no phenomenological fit (ranging from Poisson to modified
negative binomial distribution) is able to reproduce such a pattern.
They give rise either to monotonous positive ratios or to ones changing
the sign at each value of the rank.

Thus, the moment analysis is a more powerful method than the direct fits
of multiplicity distributions. The cumulants are very sensitive to any
uncertainties and require high statistics experiments at high energies.
Among different factors influencing the shapes of $H_q$-curves we
mention the cut-off of the high multiplicity tail of the distribution
(which is mainly due to limited experimental statistics), difference
between the values of moments for charged particles distributions and
those for negatives only \cite{4}, the similar difference (in model
calculations) between clusters and their decay products, and, finally,
the experimental selection criteria and error bars of $P_n$. All of them
disappear at asymptotically high energies and with statistics increased
but they can be important for any given experiment. No careful analysis
of all these factors has been done up to now. We rely on the qualitative
similarity of $H_q$-curves in all the above cases and suppose that it is
related rather to the underlying dynamics of the processes than to
varying from one experiment to another selection criteria and statistics.

Let us mention at the very end another byproduct of the analysis of the
generating functions. When the sum in eq. (\ref{1}) is cut at some final
multiplicity $N$ due to finite experimental statistics and conservation
law, the truncated generating function becomes the polynomial of $N$-th
order in $z$, i.e. it possesses $N$ complex conjugate zeros in
$z$-plane. It happens that those zeros lie near the circle of the unit
radius and at large $N$ come very close to the real axis and to the
point $z=1$ (see \cite{4}), where all the moments are calculated according to
eq.(\ref{2}). It explains why the moment analysis is so sensitive to
tiny details of the distributions. Some statistical analogies arise also
in connection with these zeros \cite{13}.

\end{document}